\documentclass[twoside]{article}

\usepackage{qic}
\usepackage{color,subfigure}
\usepackage{bm}
\usepackage{epsfig}
\usepackage[numbers]{natbib}
\usepackage{graphicx}
\usepackage{amssymb,amsmath,amsbsy,amsgen,amsfonts,amstext}
\usepackage{mathrsfs}
\usepackage{latexsym}
\usepackage{array}
\usepackage{algorithmic, algorithm}
\usepackage{epstopdf} 

\newcommand{\ket}[1]{\left\vert{#1}\right\rangle}
\newcommand{\ketbra}[2]{|#1\rangle \langle#2|}
\newcommand{\be}{\begin{equation}}
\newcommand{\ee}{\end{equation}}
\newcommand{\ba}{\begin{array}}
\newcommand{\ea}{\end{array}}
\newcommand{\bqa}{\begin{eqnarray}}
\newcommand{\eqa}{\end{eqnarray}}

\begin{document}
\setlength{\textheight}{8.0truein}    

\runninghead{Loss Tolerance with a Concatenated Graph State}
            {David A. Herrera-Mart\'{i} and Terry Rudolph}

\normalsize\textlineskip
\thispagestyle{empty}
\setcounter{page}{1}

\copyrightheading{0}{0}{2003}{000--000}

\vspace*{0.88truein}

\alphfootnote

\fpage{1}

\centerline{\bf LOSS TOLERANCE WITH A CONCATENATED GRAPH STATE}
\vspace*{0.37truein}

\centerline{\footnotesize David A. Herrera-Mart\'{i}}
\vspace*{5pt}
\centerline{\footnotesize and}
\vspace*{5pt}
\centerline{\footnotesize Terry Rudolph}
\vspace*{0.015truein}
\centerline{\footnotesize\it Controlled Quantum Dynamics Theory, Level 12 EEE, Imperial College}
\baselineskip=10pt
\centerline{\footnotesize\it London, SW7 2AZ,United Kingdom}
\vspace*{10pt}

\publisher{(received date)}{(revised date)}

\vspace*{0.21truein}

\abstracts{
A new way of addressing loss errors is introduced which combines ideas from measurement-based quantum computation and concatenated quantum codes, allowing for universal quantum computation. It is shown that for the case where qubit loss is detected upon measurement, the scheme performs well under $23\%$ loss rate. For loss rates below $10\%$ this approach performs better than the best scheme known up to date \cite{varnava2006loss}. If lost qubits are tagged prior to measurement, it can tolerate up to $50\%$ loss. The overhead per logical qubit is shown to be significantly lower than other schemes. The obtention of the threshold is entirely analytic.
}{}{}

\vspace*{10pt}

\vspace*{3pt}
\communicate{to be filled by the Editorial}

\section{Introduction}        
\noindent

Qubit loss is a common type of error that has to be dealt with effectively if we are to build a quantum processor. Loss is arguably easier to handle than unknown computational errors, since sometimes it can be seen merely as an error which can be located. Loss can happen for example as a result of fluctuations of the qubit parameters that take the state of the system out of the computational subspace, or as a result of using inefficient detectors. There exist already several works \citep{ralph2005loss, dawson2006noise, stace2009thresholds, li2010fault}, each making different assumptions and taking different error models, that combine loss protection with tolerance to unknown errors and are tailored for different architectures. However the approach to loss which is closer to the spirit of the present work was introduced by Varnava et al. \citep{varnava2006loss}, where they proved that universal quantum computation is possible even with a $50\%$ loss rate provided there are no computational errors.

We combine ideas from measurement-based quantum computation \citep{raussendorf2001one,raussendorf2003measurement} and from the traditional approach to fault tolerance \citep{gottesman2009introduction}. In particular, we use the five qubit code \citep{knill1997theory}, which is the smallest quantum error correcting code which can correct one general Pauli error \citep{Niel00}. In contrast to general Pauli errors, loss errors can be located, a property which is crucial in our construct.

We show that it is possible to achieve high levels of tolerance to qubit loss without compromising the universality of the cluster state model of computation. The whole point is to simulate a noiseless measurement pattern in a noiseless cluster state, and to this aim we encode the logical qubits using the five qubit code graph code concatenated with itself, in such a way that the logical operators to be measured will be spread across many physical qubits. We will see that the logical operators are defined as the tensor product of Pauli operators and only have support, that is, are different from the identity operator, on roughly $3/5$ of the total number of qubits, which allows for loss tolerance. A general theory for graph code concatenation can be found in refs.~\citep{grassl2009generalized, beigi2011graph}.

\section{Graph States as Error Correcting Codes}
\noindent

To define graph states \citep{Hein04} it is useful to recall the mathematical definition of a simple undirected graph $G = \{V,E\}$, where $V\subset\mathbf{N}$ are the vertices where the qubits sit. A graph state is uniquely defined as the common eigenstate of the operators $K_i = X_i\bigotimes_{\{i,j\}\in E} Z_j$,$\forall i \in V$ where $X_i$ and $Z_i$ are the Pauli matrices applied to qubit $i$.

A constructive definition would be to initially set in each vertex in the state $\ket{+}$, where $\ket{\pm}=\frac{1}{\sqrt{2}}(\ket{0}\pm\ket{1})$. The symbol $E\subset\mathbf{N}\times\mathbf{N}$ corresponds to edges connecting the qubits, representing the application of a  controlled-$Z$ operation. Indeed, it has been shown that all stabiliser codes are locally unitary equivalent to some graph state \citep{schlingemann2001quantum}. In fact graph states can be combined with non-additive classical codes to create a larger set of quantum error-correcting codes \citep{cross2008codeword}.

\subsection{A Version of the Five Qubit Code}
\noindent

\begin{figure}[h]
\centering
\subfigure[]{
\includegraphics[scale=0.25]{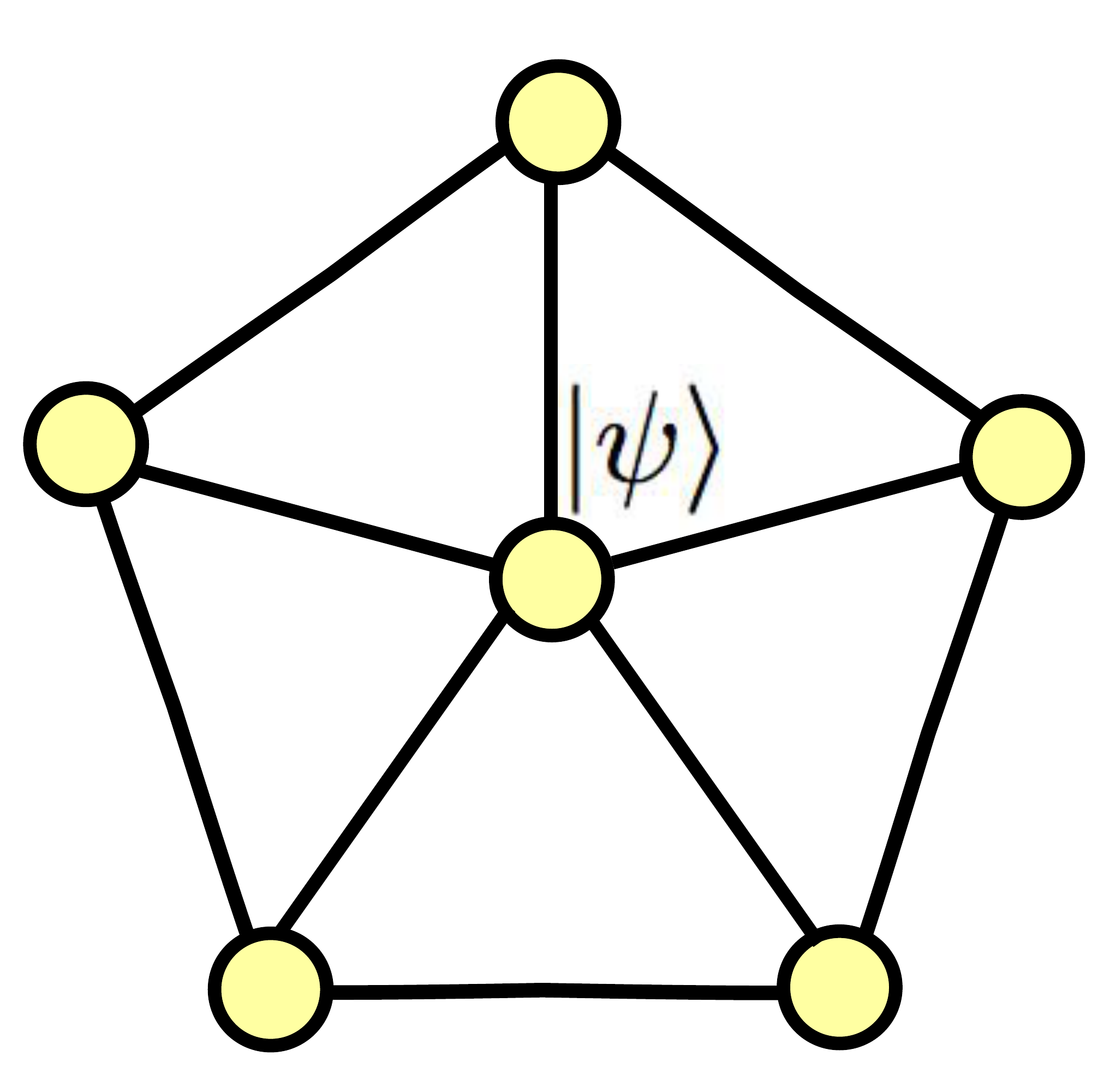}}
\subfigure[]{
\includegraphics[scale=0.6]{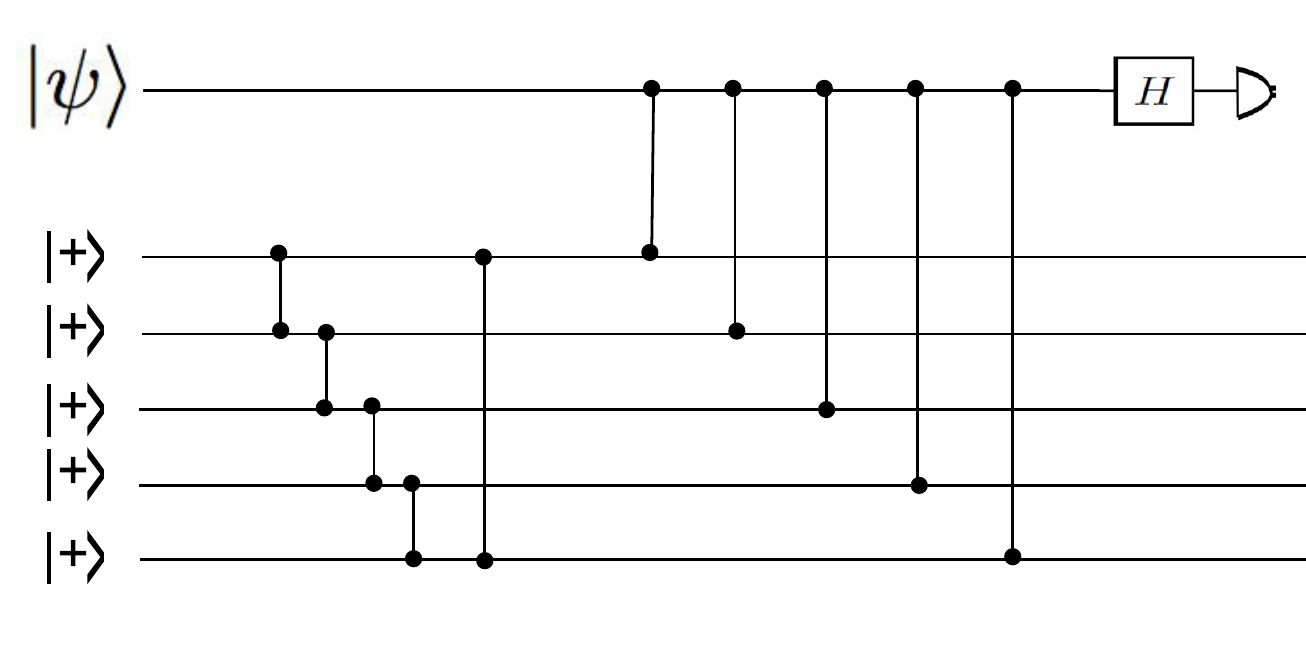}}
\fcaption{\textbf{(a)}Circles are qubits and lines between them represent the application of a $C_Z$ gate. Five qubits in a ring (a pentagon) entangled to a centre qubit is the basic building block of our scheme. The initial state is teleported into the ring via measuring the centre qubit in the X basis. \textbf{(b)} Circuit version of the encoding. The information initially contained in the centre qubit will be spread over the whole graph state after operation of the $C_Z$ gates.}
\label{encoding}
\end{figure}

Consider the circular five qubit graph state, as depicted in Fig.~\ref{encoding}(a) after removing the centre qubit. Following the definition of a graph state given above, its stabilisers are $XZIIZ$, $ZXZII$, $IZXZI$, $IIZXZ$ and $ZIIZX$. Pairwise multiplying these operators one obtains the five qubit graph code, defined as the common eigenspace of the operators $ZYYZI$, $IZYYZ$, $ZIZXX$, $XZIZX$, and $XXZIZ$, and we choose the last stabiliser $ZIIZX \equiv XXXXX \equiv \bar X$ to be the logical X operator. Note that we could have chosen any one of the original stabiliser operators. It follows that the five qubit graph state in a ring is locally unitary equivalent to the usual five qubit code initialised in the logical $\ket{+}$ state. This code saturates the singleton bound \citep{Niel00}, {\it i.e.} it is the smallest quantum code that protects against one Pauli error.  We choose $\bar Z \equiv ZZZZZ$, such that it anticommutes with $\bar X$ and commutes with the rest. This construction can be seen as a special case of \emph{codeword stabilized codes} \citep{cross2008codeword}, in which the $\bar{Z}$ is the word operator.

\subsection{Encoding and Concatenation}
\noindent

Encoding a logical qubit in the five qubit graph state can be seen as a simple teleportation circuit where we have substituted one of the qubits by the five qubit graph state, as depicted in Fig.~\ref{encoding}(b). Consider the state

\be
\ket{\psi} = \frac{1}{2}(I + e_x X + e_y Y + e_z Z)\ket{\psi},
\ee
where $e^2_x + e^2_y + e^2_z = 1$ are the amplitudes containing the information. To encode this state into the graph state, the logical controlled-$Z$ $\bar{C_Z} = \Pi^5_{i=1}C^{i}_Z$ is applied between each of the qubits in the graph and $\ket{\psi}$.

Without loss of generality we only look at the X-Z ($e_y=0$) equator of the Bloch sphere. In terms of operators, the action of the logical gate $\bar{C_Z}$ can be described as follows:

\be
\bar{C_Z}
  \left[\begin{array}{ccc}
     & I_1 &  \\
    I_5 & X &  I_2 \\
     I_4 &  & I_3 \\
  \end{array}\right]\bar{C^\dagger_Z} = \begin{array}{ccc}
     & Z_1 &  \\
    Z_5 & X &  Z_2 \\
     Z_4 &  & Z_3 \\
  \end{array},
\ee

\be
\bar{C_Z}
  \left[\begin{array}{ccc}
     & Z_1 &  \\
    X_5 & I &  I_2 \\
     Z_4 &  & I_3 \\
  \end{array}\right]\bar{C^\dagger_Z} = \begin{array}{ccc}
     & Z_1 &  \\
    X_5 & Z &  I_2 \\
     Z_4 &  & I_3 \\
  \end{array}.
\ee

The weights $e_x$ and $e_z$ of $X$ and $Z$ will, upon measurement in the $\{\ketbra{+}{+},\ketbra{-}{-}\}$ basis, be effectively stored in the amplitudes of the eigenstates of $\bar{Z}$ and $\bar{X}$, respectively. The case for $\bar{Y}$ follows through the relation $\bar{Y}=i\bar{X}\bar{Z}$. Encoding information can be seen as ``expanding" the operators acting on the original qubit onto the graph state.

To see that the logical operators have support on only three physical qubits, let us multiply them by the relevant stabilisers, as follows:

\be
 \bar{X} \equiv \begin{array}{ccc}
     & Z_1 &  \\
    X_5 & I &  I_2 \\
     Z_4 &  & I_3 \\
  \end{array} \cdot  \begin{array}{ccc}
     & Z_1 &  \\
    I_5 & I &  Y_2 \\
     Z_4 &  & Y_3 \\
  \end{array} = \begin{array}{ccc}
     & I_1 &  \\
    X_5 & I &  Y_2 \\
     I_4 &  & Y_3 \\
  \end{array}.
\ee

For $\bar{Z}$, we have:

\be
 \bar{Z} \equiv \begin{array}{ccc}
     & Z_1 &  \\
    Z_5 & I &  Z_2 \\
     Z_4 &  & Z_3 \\
  \end{array} \cdot  \begin{array}{ccc}
     & Z_1 &  \\
    I_5 & I &  Y_2 \\
     Z_4 &  & Y_3 \\
  \end{array}  = \begin{array}{ccc}
     & I_1 &  \\
    Z_5 & I &  X_2 \\
     I_4 &  & X_3 \\
  \end{array},
  \label{equation:7}
\ee

\be
 \bar{Z} \equiv \begin{array}{ccc}
     & Z_1 &  \\
    Z_5 & I &  Z_2 \\
     Z_4 &  & Z_3 \\
  \end{array} \cdot  \begin{array}{ccc}
     & Y_1 &  \\
    Y_5 & I &  Z_2 \\
     Z_4 &  & I_3 \\
  \end{array} \cdot  \begin{array}{ccc}
     & Z_1 &  \\
    Y_5 & I &  I_2 \\
     Y_4 &  & Z_3 \\
  \end{array} = \begin{array}{ccc}
     & Y_1 &  \\
    Z_5 & I &  I_2 \\
     Y_4 &  & I_3 \\
  \end{array}.
  \label{equation:8}
\ee
and all their variations derived from rotational symmetry. Thus we only need to measure three out of five qubits to retrieve the logical information. Similarly one can measure the $\bar{Y}$ operator.

\begin{figure}[h]
\centering
\subfigure[]{
\includegraphics[scale=0.16]{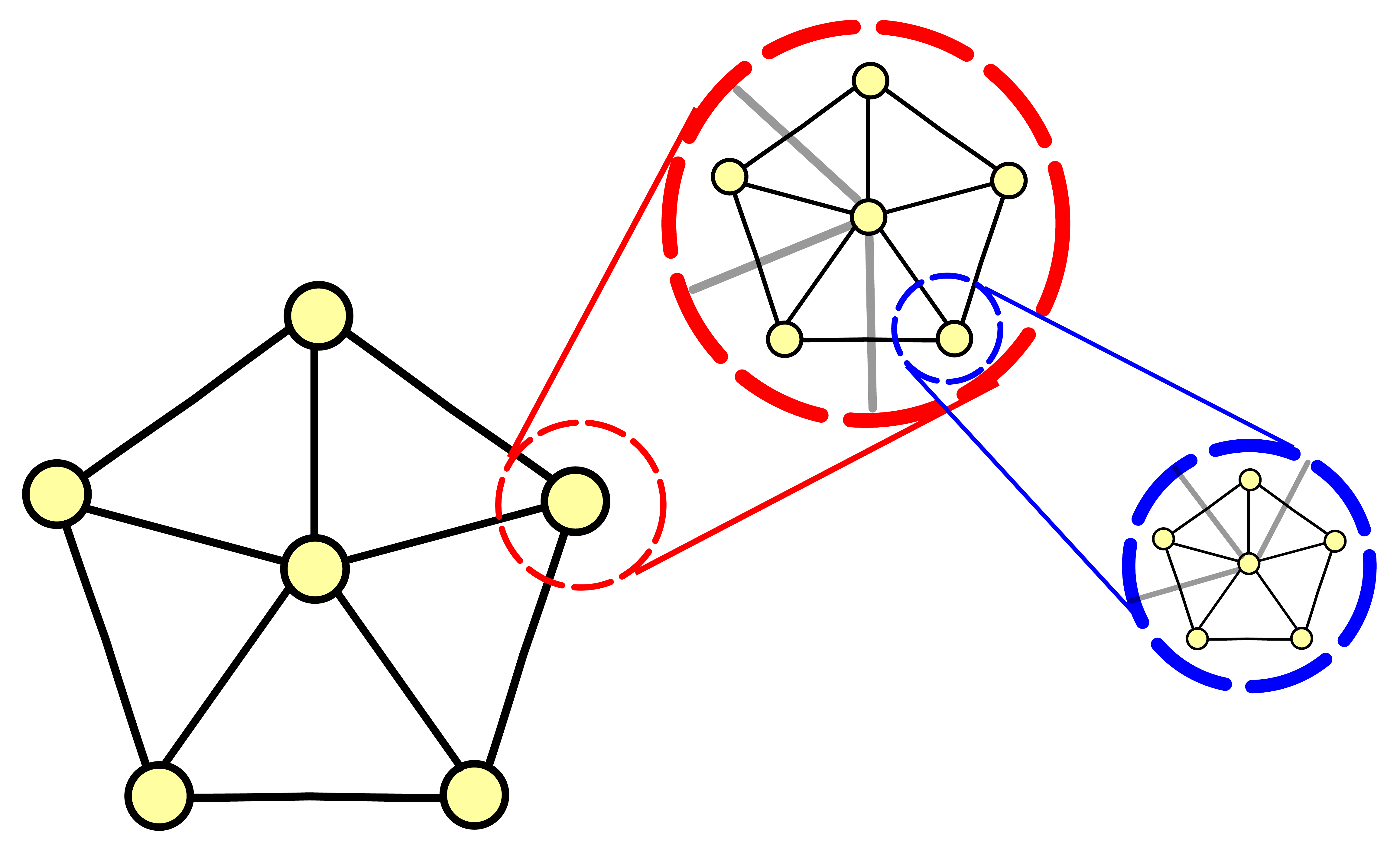}}
\subfigure[]{
\includegraphics[scale=0.16]{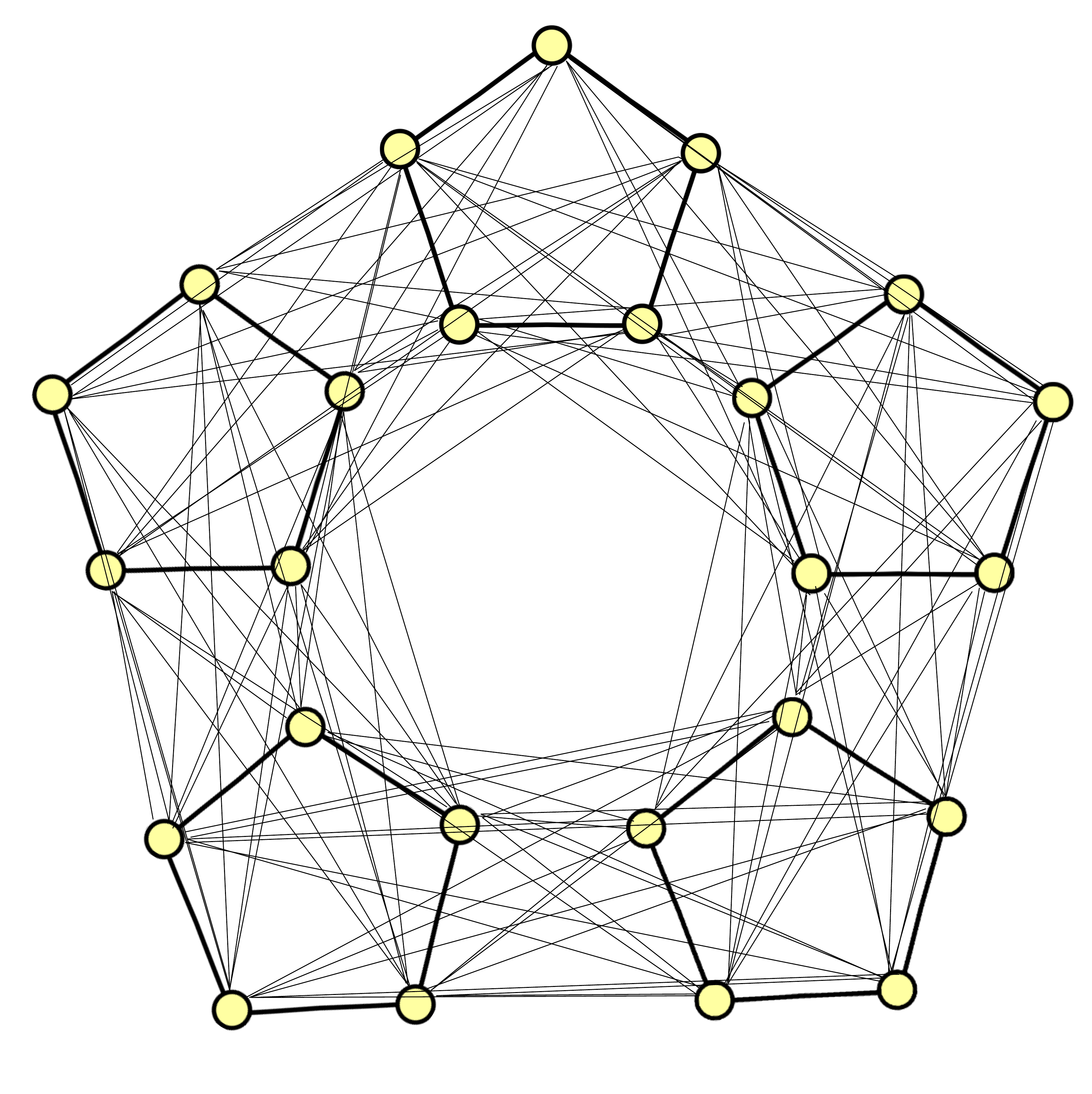}}
\fcaption{\textbf{(a)}We will stop at some level of concatenation that meets our loss tolerance requirements. \textbf{(b)}After N concatenations, the resulting logical qubit will be a constellation in which each physical qubit is entangled to $\Omega(5^{N-1})$ other qubits. Here we illustrate this for  $N=2$. It can readily be seen why tolerance to Pauli errors is not possible, since concatenation leads to exponentially many gates.}
\label{concatenation}
\end{figure}

Now, for each qubit in the pentagon, we carry out the encoding procedure explained above in this section. This is known as \emph{concatenation} (see Fig.~\ref{concatenation}(a)). To each qubit in the pentagon going out in the circuit of Fig.~\ref{encoding}, we attach a copy of that same circuit and repeat iteratively. The logical qubit is said to be at level $0$, and the qubits of the first pentagon are in level $1$. Now each qubit in level $1$ is encoded using the same code, which will create the concatenation level $2$. Iterating this procedure amounts to concatenating the code with itself $N$ times, in such a way that the last level of concatenation has $Q = 5^N$ physical qubits encoding one logical qubit (see Fig.~\ref{concatenation}(a)). It is important to realize that all previous $N-1$ levels are measured out in the $\{\ketbra{+}{+},\ketbra{-}{-}\}$ basis, without being exposed to loss, since they are just part of the definition of the code, and not part of its physical implementation.

The main idea behind code concatenation for a code correcting up to one error, is that the effective error probability at level $N$, $P^{[N]}$, decreases depends on the effective probability at level $N-1$ as:

\be
P^{[N]} = \gamma (P^{[N-1]})^2.
\ee

This can be understood as using quantum code to reduce the effective error probability of a qubit in the immediately lower level of concatenation. The probability of logical error will be the probability of two errors happening times some combinatorial constant $\gamma$, which is particular to each code and error model. This gives:

\be
P^{[N]} = \frac{(\gamma p)^{2^N}}{\gamma}.
\ee

If the physical error probability is above $1/\gamma$, the effective error probability will increase as more redundancy is added. On the contrary, if $p<1/\gamma$, there is a doubly exponential decrease of the effective error probability in the number of concatenations. For a code $[n, 1, 3]$, the number of physical qubits grows as $n^N$, {\it i.e.} exponentially in the number of concatenations, so we still have a exponential decrease of $P^{\textrm{eff}}_N$ in the number of physical qubits.

The fact that we can locate loss error leads to a slightly different way of calculating the threshold, which is explained in next section.

A general result in classical and quantum information theory \citep{Niel00} is that a code that protects against $t$ errors, can protect against $2t$ losses (loss can be regarded a localized error). It is important to stress that in our scheme we only allow for destructive measurements, that is, the qubit being measured will no longer be available to extract more information. Ultimately this is the reason why tolerance to Pauli errors cannot be integrated within this approach.

We classify loss into two broad categories: preannounced and non-preannounced. Preannounced loss happens when the absence of a qubit is detected in advance of aiming to measure it, whether the system is there or not. Non-preannounced loss means that one discovers the loss only upon measuring and not getting any ``click". This categorization is not exactly the same as the ``heralded-unheralded" division, since heralded loss means that we detect a loss at measurement and tag the location, whereas unheralded means that there is a loss that is not detected. To our understanding, this new categorization fits better in the measurement-based approach to quantum computation.

\subsection{Results for Preannounced Loss}
\noindent

In the case where we have knowledge about the location of lost qubits, a threshold for the loss probability can be derived which coincides with the theoretical maximum of $50\%$. If this maximum could be surpassed, then we would be able to copy quantum information in arbitrary basis which is precluded by the no-cloning theorem \citep{Woot82}.

We consider a physical qubit loss probability $p_L$. We show that under concatenation the effective loss probability decreases exponentially in the number of concatenations. The recurrence formula

\begin{figure}[h]
\centering
\includegraphics[scale=0.33]{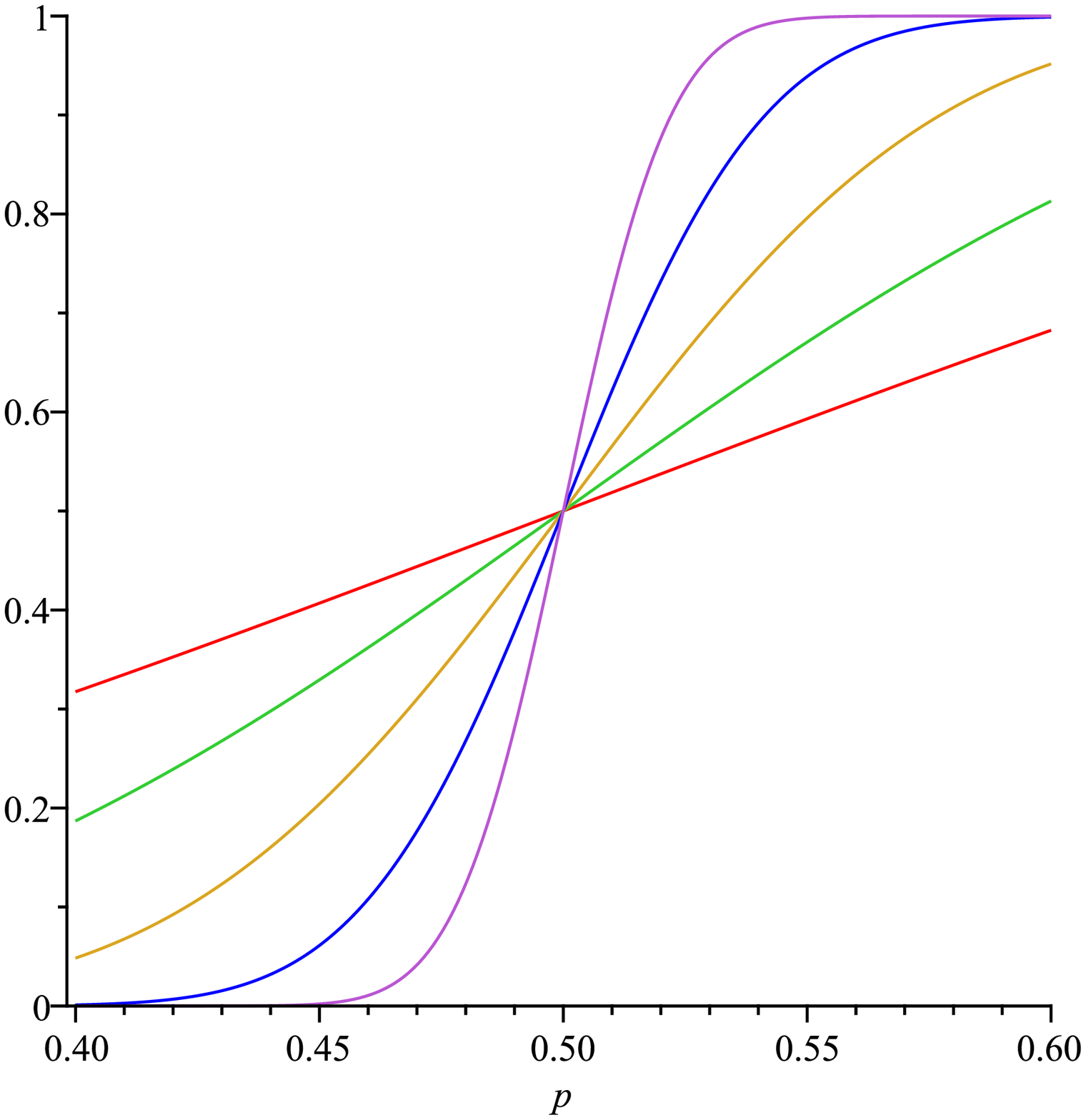}
\fcaption{Effective loss probability versus physical loss probability for preannounced loss. Different levels of concatenation, from $N=1$ ({\it cf.} $Q=5$), red,  to $N=5$ ({\it cf.} $Q=3125$), purple.}
\label{threshold}
\end{figure}

\bqa
P_L(N-1) &=& {5\choose 5}P^{5}_L(N) + {5\choose 4}P^{4}_L(N)(1 - P_L(N)) \nonumber\\
&+& {5\choose 3}P^{3}_L(N)(1 - P_L(N))^2
\eqa
gives us the effective loss probability at concatenation level $N-1$, $P_L(N-1)$, given that the loss probability at level $N$ was $P_L(N)$, with $P_L(N_{\textrm{top}}) = p_L$ at the top level. By plotting $P_L(N)$ for different $N$ it is possible come up with a recurrence fixed point corresponding to a threshold of $50 \%$, as given in Fig.~\ref{threshold}.

The recurrence formula giving the effective loss probability assumes that only the top level of concatenation is actually exposed to loss. This means that we should regard all the previous concatenation levels as levels of virtual qubits that help us visualize how to construct the code. These virtual levels are also useful in order to visualize the decoding procedure. Each virtual qubit in level $N-1$ is encoded in five qubits in level $N$, and $N_{\textrm{top}}$ corresponds to the actual physical level. Being unable to recover the information stored in any pentagon belonging to level $N$ will result in declaration of loss of the corresponding underlying qubit.

In order to gain some insight on the amount of protection given by this way of encoding for preannounced loss, Table~\ref{tab:3} illustrates the number of qubits needed in to make the effective loss probability $P_L(N)\approx10^{-8}$ or below:

\begin{table}[!h]
\tcaption{Number of physical qubits $Q^V$ (for the trees approach) and $Q$ (for this proposal) used to achieve an effective loss probability $P_L(N)\approx10^{-8}$ or below.}
\centering
\begin{tabular}{|c||c|c|c|}
\hline
     & $p_L=0.2$ & $p_L=0.3$ & $p_L=0.4$ \\
\hline
\hline
$Q^V$ & 22188 & $2.3\times10^5$ & $7.6\times10^6$ \\
\hline
  $Q$ & 125 & 625 & 3125 \\
\hline
\end{tabular}.
\label{tab:3}
\end{table}

Here we have compared our results with the amount of qubits needed in \citep{varnava2006loss}. We stress that in our case, these values are valid only when loss is preannounced, as opposed to \citep{varnava2006loss} where which particular qubits have been lost need not be known beforehand. We nevertheless include this table to show that resources would be dramatically reduced if one could tag lost qubits prior to measurement, such as may be relevant for atoms in optical lattices.

\subsection{Results for Non-preannounced Loss}
\noindent

The bound of $50\%$ is achieved when one knows whether the qubit is there or not. If, as we now consider, one discovers a loss while performing a computational measurement, there is a chance that the measurement pattern chosen prior to discovering the loss will be unavoidably broken and the information lost. If this happens, then one declares a loss, which can be handled in the same way as loss in the immediately lower concatenation level.

Since each logical operator can be written in two different (commuting) ways, one can try to measure both ways at the same time and declare a loss whenever either two losses break both of them, or when a loss breaks one of them but it is impossible go back and to measure the other one. The details of a measurement strategy that maximises the probability of measuring \emph{at least} one version of the logical operator are contained in Decision Tree~\ref{decision}. This measurement strategy gives rise to a new recurrence formula that can be used to obtain the new threshold, which is considerably lowered to about $23\%$ percent. However, it comparable to thresholds obtained for other architectures \citep{stace2009thresholds, stace2010error}.

\begin{figure}[h]
\centering
\includegraphics[scale=0.33]{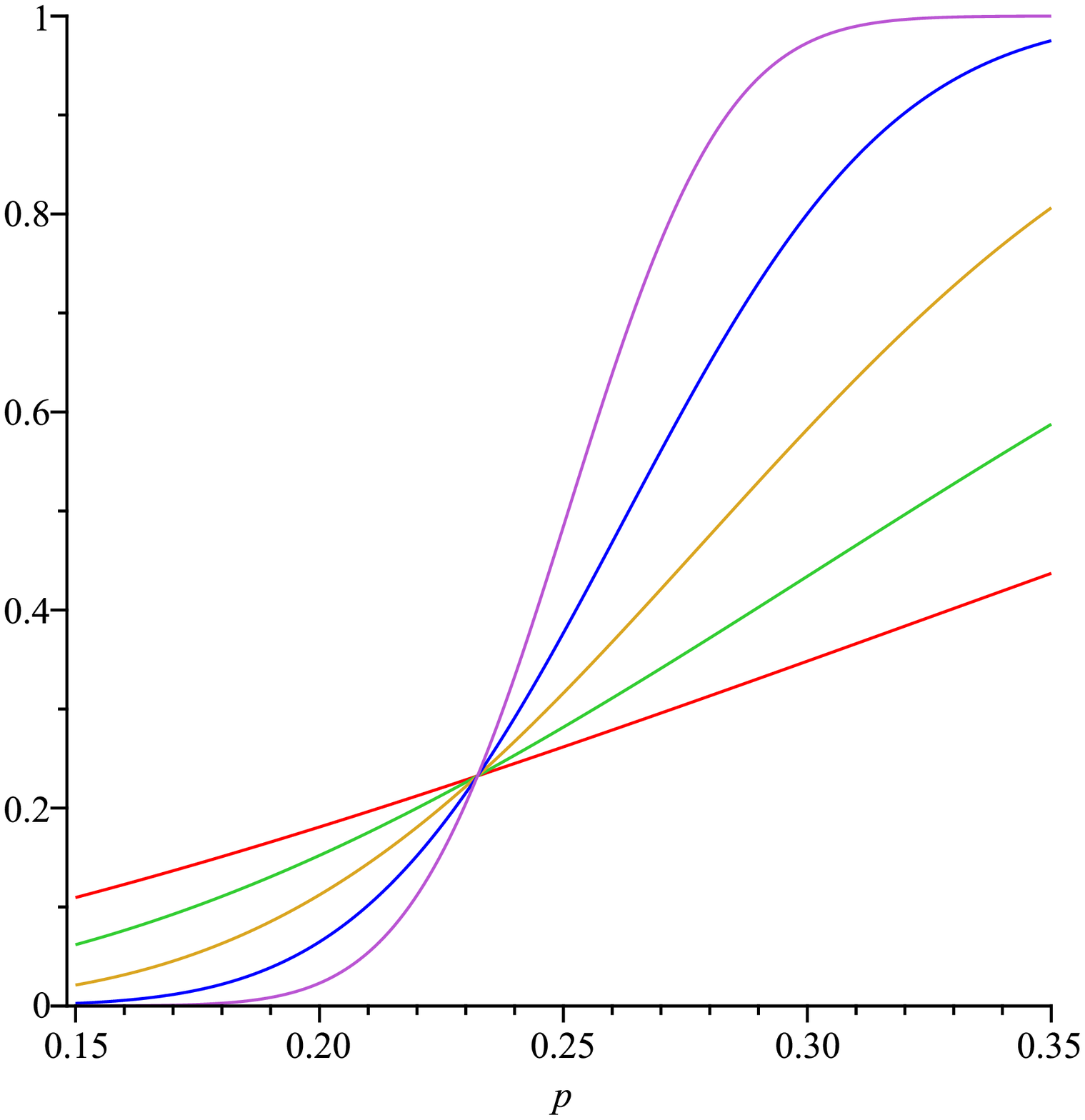}
\fcaption{Effective loss probability versus physical loss probability for non-preannounced loss. Different levels of concatenation, from $N=1$ ({\it cf.} $Q=5$), red,  to $N=5$ ({\it cf.} $Q=3125$), purple.}
\label{nonpreannounced}
\end{figure}

\subsubsection{Measurement strategy for non-preannounced loss}
\noindent

The decision tree for a $\bar{Z}$ measurement is given in the Decision Tree~\ref{decision}, written in pseudocode. The notation ``$jA$" means ``qubit $j$ is measured in the $A$ basis". 

The algorithm reduces to an attempt to measure a correlation conforming a logical operator. A logical operator can be written in different forms, as shown for example in equations~\ref{equation:7} and \ref{equation:8}. Upon discovering a loss error, the operator being measured may or may not be retrieved, depending on whether changing the basis in which the remaining qubits are measured allows to switch from one correlation to another.

\vspace{5mm}
\floatname{algorithm}{Decision Tree}
\begin{algorithm}[!h]
\caption{ for a $\bar{Z}$ measurement}
\algsetup{indent=3em}
\begin{algorithmic}[1]
\IF{1X}
 	\IF{2Z}
		\IF{5Z} \STATE SUCCESS
		\ELSIF{3Y}
			\IF{5Y} \STATE SUCCESS
			\ELSE \STATE FAILURE
			\ENDIF
		\ELSE \STATE FAILURE
		\ENDIF
	\ELSIF{3Y}
		\IF{5Y} \STATE SUCCESS
		\ELSE \STATE FAILURE
		\ENDIF
	\ELSE \STATE FAILURE
	\ENDIF
\ELSIF{2X}
	\IF{4Y}
		\IF{5Y} \STATE SUCCESS
		\ELSE \STATE FAILURE
		\ENDIF
	\ELSE \STATE FAILURE
	\ENDIF
\ELSIF{4X}
		\IF{3Z}
			\IF{5Z} \STATE SUCCESS
			\ELSE \STATE FAILURE
			\ENDIF
		\ELSE \STATE FAILURE
		\ENDIF
\ELSE \STATE FAILURE
\ENDIF
\end{algorithmic}
\label{decision}
\end{algorithm}

There are similar decision trees for $\bar{X}$ and $\bar{Y}$ measurements. Using these pseudocodes recursively, {\it i.e.} at each level of concatenation, will give us a measurement strategy in the limit of many concatenations. Adding probabilities for success and failure yields the curve in Fig.~\ref{nonpreannounced}.

\section{Loss-tolerant Universal Quantum Computation}
\noindent

To see how universality is achieved, keep in mind that a logical graph state underlies all encodings. We assume this graph state is two-dimensional, so that controlled-Z gates are naturally embedded in the encoding. As shown in Fig.~\ref{encoding} (b), encoding can be seen as entangling the operators of the virtual qubit with the logical operators living in the pentagons. Imagine we have two such virtual qubits entangled with a $C_Z$ gate. It is trivial to see how this translates into entanglement between the logical operators of their respective encodings.

\be
\begin{array}{c}
  \bar I \\  I \\  I \\  \bar X
\end{array}\begin{array}{c}
  \bar I \\  Z \\  X \\  \bar I
\end{array}\begin{array}{c}
  \bar I \\  X \\  Z \\  \bar I
\end{array}\begin{array}{c}
  \bar X \\  I \\  I \\  \bar I
\end{array}\Rightarrow\begin{array}{c}
  \bar I \\  I \\  Z \\  \bar X
\end{array}\begin{array}{c}
  \bar I \\  Z \\  X \\  \bar Z
\end{array}\begin{array}{c}
  \bar Z \\  X \\  Z \\  \bar I
\end{array}\begin{array}{c}
  \bar X \\  Z \\  I \\  \bar I
\end{array}\Rightarrow\begin{array}{c}
  \bar X \\  I \\  X \\  \bar Z
\end{array}\begin{array}{c}
  \bar I \\  I \\  X \\  \bar I
\end{array}\begin{array}{c}
  \bar I \\  X \\  I \\  \bar I
\end{array}\begin{array}{c}
  \bar Z \\  X \\  I \\  \bar X
\end{array},
\ee
where we start off with two virtual qubits (centre) in an entangled state. The first arrow represents the encoding of the virtual qubits ({\it i.e.} entangle them with logical operators), and the second arrow represents measurement of the virtual qubits in the X basis. This shows that the logical operators are entangled via a logical $C_Z$ gate.

Unfortunately, only measurements in the X, Y and Z basis can be done in a loss-tolerant fashion. This prevents us from doing single qubit gates with the usual prescription ({\it i.e.} steer a unitary by measuring in angles given by it's Euler decomposition). This can be overcome by introducing the additional set of gates  $\exp(i\frac{1}{8})Z$, $\exp(i\frac{1}{4})Z$ and $\exp(i\frac{1}{4})X$. These gates can be realized fault-tolerantly using a special type of error-free states known as \emph{magic states} \citep{bravyi2005universal}. Magic states can be distilled using from a reservoir of not-too-noisy ancillas using only measurements in the X basis and $C_Z$ gates.

\begin{figure}[h]
\centering
\includegraphics[scale=0.4]{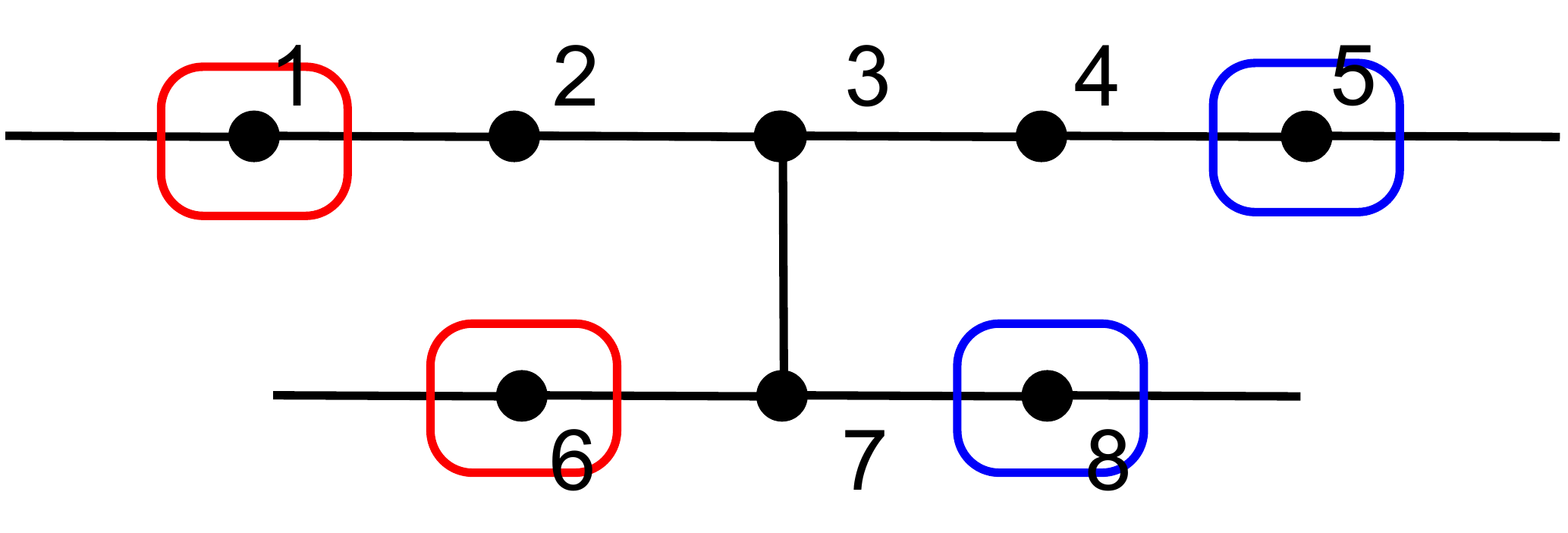}
\fcaption{This is how a $C_X$ gate would look like in the virtual graph state. Red squares denote input states and blue squares denote output states. All qubits except the blue ones are measured in the X basis.}
\label{cnot}
\end{figure}

The correlations defining the $C_X$ gate of Fig.~\ref{cnot} are:

\bqa
& &X_1I_2X_3I_4X_5I_7X_8, \\
& &Z_1X_2I_3X_4Z_5, \\
& &I_3X_4Z_5Z_6X_7Z_8, \\
& &X_6I_7X_8.
\eqa

Measuring all qubits of Fig.~\ref{cnot} in the X basis will enact a $C_X$ gate. Hadamard gates are also straightforward to achieve. This is clear if we again have a look at the virtual graph state, since measuring in the X basis a qubit in state $\ket{\psi}$ will steer the next qubit in the graph into the state $X^m H \ket{\psi}$, where $m$ is the outcome of the measurement.

\subsection{Overhead}
\noindent

The basic principle of fault tolerance using concatenated codes is that, whenever the physical qubit error probability is below the threshold, the effective error probability decreases exponentially with the number of physical qubits. However, as one gets closer to the threshold, the resources needed to maintain a given effective loss probability $P_L$ increase very fast. The effective loss probability $P_L$ as a function of the overhead $Q_\textrm{pre} = 5^N$ and the physical loss probability $p_L$ is summarized in Tables~\ref{tab:4} and \ref{tab:5} for preannounced and non-preannounced loss, respectively.

\begin{table}[!h]
\tcaption{Effective loss probability as a function of the number of qubits $Q_\textrm{pre}$ and the physical loss probability $p_L$, for the case of preannounced loss.}
\centering
\begin{tabular}{|c||c|c|c|}
\hline
  $Q_\textrm{pre}$ & $p_L=0.4$ & $p_L=0.3$ & $p_L=0.2$\\
\hline
\hline
  5 & 0.317 & 0.163 & 0.058 \\
  25 & 0.187 & 0.033 & 0.002\\
  125 & 0.048 & $3.6\times10^{-4}$ & $5.6\times10^{-8}$\\
  625 & 0.001 & $4.5\times10^{-10}$ & $1.8\times10^{-21}$\\
  3125 & $1.5\times10^{-8}$ & $9.1\times10^{-28}$ & $5.5\times10^{-62}$ \\
\hline
\end{tabular},
\label{tab:4}
\end{table}

\begin{table}[!h]
\tcaption{Effective loss probability as a function of the number of qubits $Q_\textrm{non-pre}$ and the physical loss probability $p_L$, for the case of non-preannounced loss.}
\centering
\begin{tabular}{|c||c|c|c|}
\hline
  $Q_{\textrm{non-pre}}$ & $p_L=0.15$ & $p_L=0.1$ & $p_L=0.05$\\
\hline
\hline
  5 & 0.110 & 0.052 & 0.014 \\
  25 & 0.062 & 0.015 & 0.001\\
  125 & 0.021 & 0.001 & $8.0\times10^{-6}$\\
  625 & 0.002 & $1.1\times10^{-5}$ & $3.8\times10^{-10}$\\
  3125 & $4.1\times10^{-5}$ & $7.7\times10^{-10}$ & $8.9\times10^{-19}$ \\
\hline
\end{tabular}.
\label{tab:5}
\end{table}

\subsection{Comparison with Tree Codes}
\noindent

We show now that, for low enough $p_L$, this approach necessitates less resources than a previous scheme which attains also the highest threshold achievable. It was introduced by M. Varnava et al. in \citep{varnava2006loss} and offers protection even agains non-preannounced loss for $p_L$ up to $50\%$. The graph states they introduce consist of ``trees" of qubits were at each level of the tree structure, the branching parameter is potentially different from the others. Results were obtained by exhaustive search over the branching parameter.

\begin{figure}[h]
\centering
\includegraphics[scale=0.3]{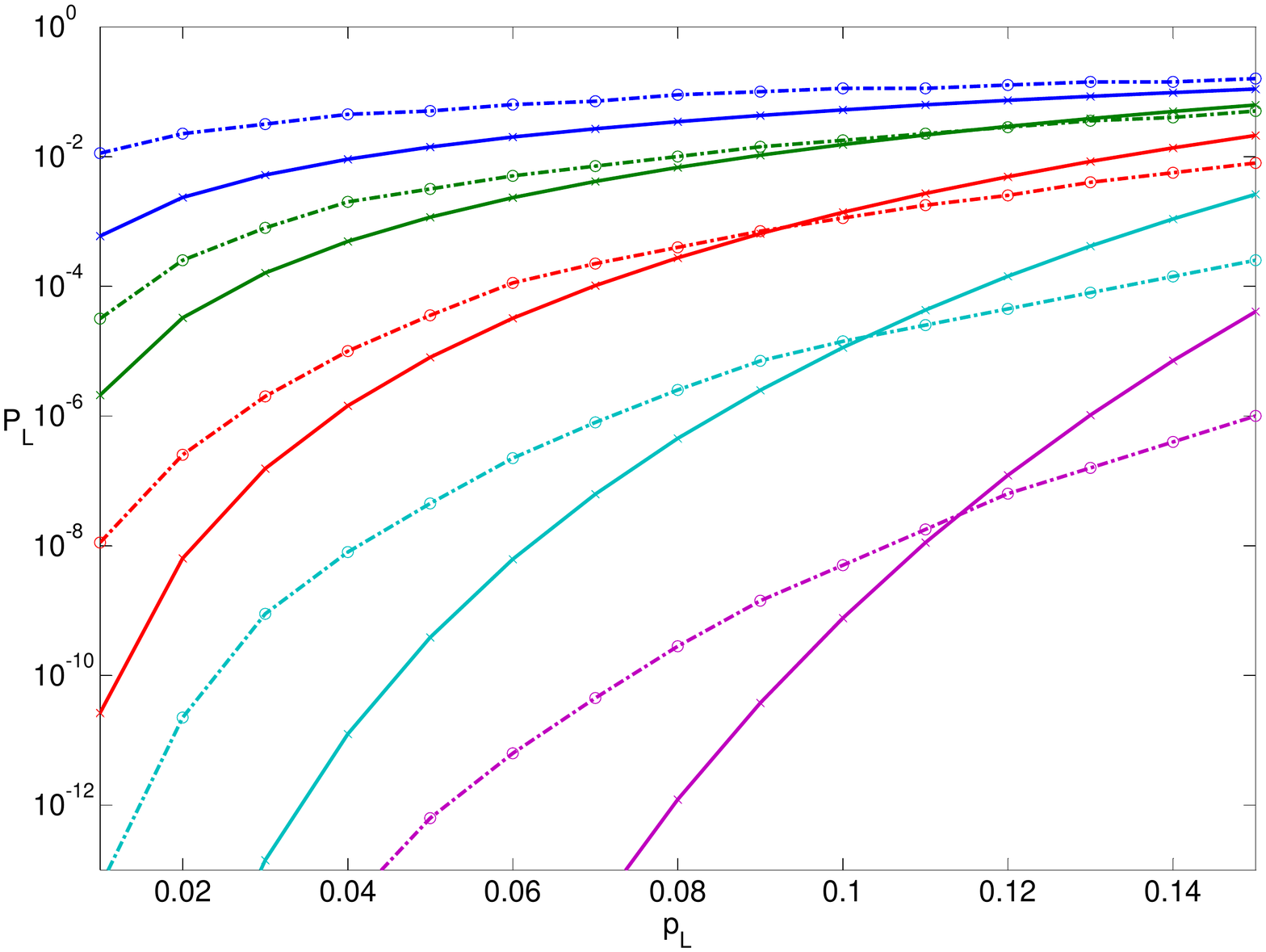}
\fcaption{Comparison of the effective loss probability $P_L$ as a function of the physical loss probability $p_L$, for both the current approach (solid lines) and the the tree codes (dashed lines). The colors correspond to different numbers of redundant physical qubits $Q$. Blue lines correspond to $Q\approx5$ qubits, green to $Q\approx25$, red to $Q\approx125$, cyan to $Q\approx625$ and purple to $Q\approx3125$. Clearly, the current approach yields better protection for low loss probabilities.}
\label{trees_pentagons}
\end{figure}

We investigated whether there is a regime in which the approach of concatenated pentagons performs better, in terms of overhead needed, than the tree approach. We found that this is the case for a non-preannounced loss probability below $\sim10\%$, as can be inferred from Fig.~\ref{trees_pentagons}.

To summarize, this scheme requires to our knowledge the least overhead for preannounced loss. For non-preannounced loss, it performs better than the trees below $\sim10\%$, which is to date the scheme with the best performance for non-preannounced loss.

\section{Conclusions and Outlook}
\noindent

We have introduced a new way to fight loss errors and shown that it allows for universal quantum computation. Comparing this with Varnava's results, we see that this new approach demands significantly less overhead for loss rates below $\sim10\%$. For preannounced loss, this scheme saturates the upper bound given by the no-cloning theorem with very few resources.

However, it seems difficult to combine it with tolerance to computational errors since $\Omega(5^N)$ gates are needed in order to provide protection at N levels of concatenation. This, even for very low unknown error probabilities, will effectively randomize the encoded qubit, since it only takes one wrong measurement to change the parity of the logical qubit. There are, however, architechtures such as optical quantum computing, where qubits are in an essentially zero temperature environment, so loss is a far more significant error than Pauli errors.

We analyzed other codes, such as the four qubit code and the seven qubit code, but their performance was significantly worse than the five qubit code, so we didn't continue in that direction. We conjecture that this is directly related to the singleton bound, since the five qubit code achieves error correction with minimal resources. On the other hand, the advantage of the five qubit code also comes from the fact that it is more symmetric than the Steane code. An interesting open question is whether larger rings of qubits can give rise to the same loss protection while increasing the number of encoded qubits.

Also, there is an exciting connection between the erasure channel and access structures \citep{markham2008graph}. The concatenated nature of this scheme gives rise to a fractal network topology, which potentially arises in several networking scenarios. A more in depth study of this connection would therefore be of interest.

\section{Acknowledgments}
\noindent
The authors are indebted with Mark Tame, Damian Markham and Mike Varnava for inspiring conversations. DHM would like to acknowledge support from the IDEA League program. This work is supported by UK Engineering and Physical Sciences Research Council.

\bibliographystyle{unsrt}
\bibliography{biblio}

\begin{thebibliography}{10}

\bibitem{varnava2006loss}
{Varnava, M. and Browne, D.E. and Rudolph, T.}
\newblock {Loss Tolerance in One-way Quantum Computation via Counterfactual
  Error Correction}.
\newblock {\em {Phys. Rev. Lett.}}, {97}({12}):{120501}, {2006}.

\bibitem{ralph2005loss}
{Ralph, T.C. and Hayes, AJF and Gilchrist, A.}
\newblock {Loss-tolerant Optical Qubits}.
\newblock {\em {Phys. Rev. Lett.}}, {95}({10}):{100501}, {2005}.

\bibitem{dawson2006noise}
{Dawson, C.M. and Haselgrove, H.L. and Nielsen, M.A.}
\newblock {Noise Thresholds for Optical Quantum Computers}.
\newblock {\em {Phys. Rev. Lett.}}, {96}({2}):{20501}, {2006}.

\bibitem{stace2009thresholds}
{Stace, T.M. and Barrett, S.D. and Doherty, A.C.}
\newblock {Thresholds for Topological Codes in the Presence of Loss}.
\newblock {\em {Phys. Rev. Lett.}}, {102}({20}):{200501}, {2009}.

\bibitem{li2010fault}
{Li, Y. and Barrett, S.D. and Stace, T.M. and Benjamin, S.C.}
\newblock {Fault Tolerant Quantum Computation with Nondeterministic Gates}.
\newblock {\em {Phys. Rev. Lett.}}, {105}({25}):{250502}, {2010}.

\bibitem{herrera2010photonic}
D.A. {Herrera-Mart{\'i}}, A.G. Fowler, D.~Jennings, and T.~Rudolph.
\newblock {Photonic Implementation for the Topological Cluster-state Quantum
  Computer}.
\newblock {\em {Phys. Rev. A}}, {82}({3}):{032332}, {2010}

\bibitem{raussendorf2001one}
{Raussendorf, R. and Briegel, H. J.}
\newblock {A One-way Quantum Computer}.
\newblock {\em {Phys. Rev. Lett.}}, {86}({22}):{5188--5191}, {2001}.

\bibitem{raussendorf2003measurement}
{Raussendorf, R. and Browne, D.E. and Briegel, H. J.}
\newblock {Measurement-based Quantum Computation on Cluster States}.
\newblock {\em {Phys. Rev. A}}, {68}({2}):{022312}, {2003}.

\bibitem{gottesman2009introduction}
{Gottesman, D.}
\newblock {An Introduction to Quantum Error Correction and Fault-tolerant
  Quantum Computation}.
\newblock pages {13--58}, {2009}.

\bibitem{knill1997theory}
{Knill, E. and Laflamme, R.}
\newblock {Theory of Quantum Error-correcting Codes}.
\newblock {\em {Phys. Rev. A}}, {55}({2}):{900}, {1997}.

\bibitem{Niel00}
{M. A. Nielsen and I. L. Chuang}.
\newblock {\em {Quantum Computation and Quantum Information}}.
\newblock Cambridge University Press.

\bibitem{grassl2009generalized}
{Grassl, M. and Shor, P. and Smith, G. and Smolin, J. and Zeng, B.}
\newblock {Generalized Concatenated Quantum Codes}.
\newblock {\em {Phys. Rev. A}}, {79}({5}):{050306}, {2009}.

\bibitem{beigi2011graph}
{Beigi, S. and Chuang, I. and Grassl, M. and Shor, P. and Zeng, B.}
\newblock {Graph Concatenation for Quantum Codes}.
\newblock {\em { J. of Mathematical Physics}}, {52}:{022201}, {2011}.

\bibitem{Hein04}
{M. Hein and J. Eisert and H.- J. Briegel}.
\newblock {Multi-party Entanglement in Graph States}.
\newblock {\em {Phys. Rev. A}}, {69}:{062311}, {2004}.
\newblock {quant-ph/0307130}.

\bibitem{schlingemann2001quantum}
{Schlingemann, D. and Werner, R.F.}
\newblock {Quantum Error-correcting Codes Associated with Graphs}.
\newblock {\em {Phys. Rev. A}}, {65}({1}):{012308}, {2001}.

\bibitem{cross2008codeword}
{Cross, A. and Smith, G. and Smolin, J.A. and Zeng, B.}
\newblock {Codeword Stabilized Quantum Codes}.
\newblock pages {364--368}, {2008}.

\bibitem{Woot82}
{W. K. Wootters and W. H. Zurek}.
\newblock {A single quantum cannot be cloned}.
\newblock {\em {Nature}}, {299}:{802--803}, {1982}.

\bibitem{stace2010error}
{Stace, T.M. and Barrett, S.D.}
\newblock {Error Correction and Degeneracy in Surface Codes Suffering Loss}.
\newblock {\em {Phys. Rev. A}}, {81}({2}):{22317}, {2010}.

\bibitem{bravyi2005universal}
{Bravyi, S. and Kitaev, A.}
\newblock {Universal Quantum Computation with Ideal Clifford Gates and Noisy
  Ancillas}.
\newblock {\em {Phys. Rev. A}}, {71}({2}):{22316}, {2005}.

\bibitem{markham2008graph}
D.~Markham and B.C. Sanders.
\newblock Graph states for quantum secret sharing.
\newblock {\em Physical Review A}, 78(4):042309, 2008.

\end{thebibliography}

\end{document}